\documentclass[prl,twocolumn,superscriptaddress]{revtex4}
\usepackage{amssymb,amsmath,amsthm,graphicx}
\usepackage{latexsym}
\usepackage{bm}
\usepackage[]{hyperref}
\allowdisplaybreaks

\pdfoutput=1

\begin{document}

\title{Localised $\bf{AdS_5\times S^5}$ Black Holes}

\author{ \'Oscar J. C. Dias}
\email{ojcd1r13@soton.ac.uk}
\affiliation{STAG research centre and Mathematical Sciences, University of Southampton, UK} 
\author{Jorge E. Santos}
\email{jss55@cam.ac.uk}
\affiliation{Department of Applied Mathematics and Theoretical Physics, University of Cambridge, Wilberforce Road, Cambridge CB3 0WA, UK} 
\author{Benson Way}
\email{bw356@cam.ac.uk}
\affiliation{Department of Applied Mathematics and Theoretical Physics, University of Cambridge, Wilberforce Road, Cambridge CB3 0WA, UK} 

\begin{abstract}
We numerically construct asymptotically global $\mathrm{AdS}_5\times \mathrm{S}^5$ black holes that are localised on the $\mathrm{S}^5$. These are solutions to type IIB supergravity with $\mathrm S^8$ horizon topology that dominate the theory in the microcanonical ensemble at small energies. At higher energies, there is a first-order phase transition to $\mathrm{AdS}_5$-Schwarzschild$\times \mathrm{S}^5$. By the AdS/CFT correspondence, this transition is dual to spontaneously breaking the $SO(6)$ R-symmetry of $\mathcal N=4$ super Yang-Mills down to $SO(5)$. We extrapolate the location of this phase transition and compute the expectation value of the resulting scalar operators in the low energy phase. 
\end{abstract}

\maketitle

{\bf 1.~Introduction --} Since its discovery, the duality between type IIB supergravity on $\mathrm{AdS}_5\times \mathrm{S}^5$ and $\mathcal N=4$ supersymmetric Yang-Mills (SYM) with large $N$ gauge group $SU(N)$ and large t'Hooft coupling remains our best understood example of a gauge/gravity duality \cite{Maldacena:1997re,Gubser:1998bc,Witten:1998qj,Aharony:1999ti}.  However, the properties of low energy states in these theories are only understood heuristically \footnote{We only consider the strict $N\to+\infty$ limit, where small energy means $E=\alpha N^2$ with $\alpha$ small.}.  

Consider type IIB supergravity on \emph{global} $\mathrm{AdS}_5\times \mathrm{S}^5$. The asymptotic region of global $\mathrm{AdS}_5$ is dual to a gauge theory background spacetime that is conformal to $\mathbb R^{(t)}\times S^3$.  In the gravity theory, the dominant low energy states are expected to be black holes with horizon topology $\mathrm S^8$ that have `localised' on the $\mathrm{S}^5$.  That is, the horizon covers one of the poles of the $\mathrm{S}^5$. These black holes would then describe thermal states where the $SO(6)$ R-symmetry of the scalar sector of SYM has been spontaneously broken.  

The existence of a symmetry-breaking phase transition can be inferred from a Gregory-Laflamme instability whereby small $\mathrm{AdS}_5$-Schwarzschild$\times \mathrm{S}^5$ ($\mathrm{AdSSchw}_5\times \mathrm{S}^5$) black holes \cite{Hawking:1982dh,Witten:1998zw} become unstable to deformations of the $\mathrm{S}^5$ due to a separation of horizon length scales \cite{Gregory:1993vy,Martinec:strings98,Banks:1998dd,Peet:1998cr,Hubeny:2002xn,Dias:2015pda,Buchel:2015gxa,Dias:2015nua}. However, only recently has it been found that the phase transition is first order \cite{Dias:2015pda}. The location of the phase transition and the expectation value of the scalar operators remain unknown.

In this manuscript, we perform a numerical construction of these localised black holes in type IIB supergravity.  We demonstrate that these solutions dominate the microcanonical ensemble over $\mathrm{AdSSchw_5}\times \mathrm{S}^5$ at small energies, extrapolate the location of the phase transition, and compute the expectation value of the scalar operators in the dual field theory. 

{\bf 2.~Numerical Construction --} The minimal field content in type IIB supergravity that can be asymptotically AdS$_5\times \mathrm{S}^5$ consists of a metric  $g$ and a Ramond-Ramond self-dual 5-form $F_{(5)}=\mathrm{d} C_{(4)}$. In this sector of the theory, the classical equations of motion are
\begin{subequations}\label{eq:eom}
\begin{align}
        E_{MN}\equiv R_{MN} - \frac{1}{48} F_{MPQRS} F_N{}^{PQRS}&=0\;,\label{eq:einstein}\\
        \nabla_M F^{MPQRS}&=0\;,\\
        F_{(5)}&=\star F_{(5)}\;.
\end{align}
\end{subequations}
We seek static, topologically $\mathrm S^8$ black hole solutions that are asymptotically $\mathrm{AdS}_5\times \mathrm{S}^5$.    Gravitational intuition suggests that the most symmetric of such black holes will have the largest entropy. These have $\mathbb R^{(t)}\times SO(4)\times SO(5)$ symmetry, where the full $SO(4)$ symmetry of $\mathrm{AdS}_5$ and the largest subgroup of $SO(6)$ are preserved. 

We use the DeTurck method \cite{Headrick:2009pv,Figueras:2011va} (see \cite{Dias:2015nua} for a review). The method requires the choice of a reference metric $\overline g$ from which one obtains the DeTurck vector 
\begin{equation}
\xi^M \equiv g^{PQ}[\Gamma^M_{PQ}-\overline{\Gamma}^M_{PQ}]\;,
\end{equation}
where $\Gamma^M_{PQ}$ and $\overline{\Gamma}^M_{PQ}$ define the Levi-Civita connections for $g$ and $\bar g$, respectively. One then modifies the Einstein equation \eqref{eq:einstein} to the equation 
\begin{equation}\label{eq:deturck}
E_{MN}-\nabla_{(M}\xi_{N)}=0\;.
\end{equation}
This equation, unlike \eqref{eq:einstein}, will yield a set of PDEs that are elliptic in character. But after solving these PDEs, one must verify that $\xi^M=0$ in order to confirm that a solution to \eqref{eq:einstein} has indeed been found.  The local uniqueness property of elliptic equations guarantees that solutions with $\xi^M=0$ are distinguishable from those with $\xi^M\neq0$.  Incidentally, the condition $\xi^M=0$ would also fix all coordinate freedom in the metric. 

Now let us describe our reference metric and ansatz.  The reference metric $\bar g$ must contain the same symmetries and causal structure as the desired solution.  Consider
\begin{subequations}\label{eq:ansatzxy}
\begin{align}
       \mathrm ds^2&= \frac{L^2}{\left(1-y^2\right)^2} {\biggl [} - \frac{1}{L^2}\,H_1 f_1 {\mathrm d}t^2+\nonumber\\
       &\quad\qquad+H_2 \left(\frac{4 f_2}{2-y^2}{\mathrm d}y^2+  y^2 \left(2-y^2\right) f_3 \, {\mathrm d}\Omega_3^2 \right)\biggr ]+  \nonumber \\
& \quad  + L^2 H_2 {\biggl [} \frac{16}{2-x^2} f_4 \left({\mathrm d}x+ f_6 {\mathrm d}y\right)^2 
+\nonumber\\
 &\quad\qquad\qquad+4 x^2 \left(2-x^2\right) \left(1-x^2\right)^2 f_5 \, {\mathrm d}\Omega_4^2  \biggr ]\;,\\
          C_{(4)}&=L^3 \frac{y^4 \left(2-y^2\right)^2}{\sqrt{2} \left(1-y^2\right)^4} H_1 f_7 \,{\mathrm d}t  \wedge {\mathrm d}S_{(3)}+L^4 W \,\mathrm dS_{(4)} \;,
\end{align}
\end{subequations}
where $L$ is the curvature scale of the boundary $S^3$, $f_1,\ldots,f_7$ are unknown functions of $x$ and $y$, and $H_1$, $H_2$ are known functions which we will describe shortly.  The function $W$ can be algebraically eliminated from the equations of motion, and can be computed after $f_1,\ldots,f_7$ are known \footnote{$W$ can be eliminated from the metric and the field strength $F_{(5)}$, so it is not needed to extract physical quantities.}.   This is a general ansatz consistent with the required symmetries.  Further note that if we set $f_6=0$, and $f_i=1$ for $i\neq 6$, as well as $H_1=H_2=1$, we recover global $\mathrm{AdS}_5\times \mathrm{S}^5$ \cite{Witten:1998zw} .  

Accordingly, we will define our reference metric by setting $f_6=0$, and $f_i=1$ for $i\neq 6$. It remains for us to supply $H_1$ and $H_2$ to fully specify the reference metric.  These must approach $H_1=H_2=1$ at $y=1$ in order to recover global $\mathrm{AdS}_5\times \mathrm{S}^5$ asymptotically.  They must also be chosen so that the reference metric describes a regular $\mathrm S^8$ black hole.  To aid in finding the solution, we would like the geometry near the horizon to be that of 10-dimensional asymptotically flat Schwarzschild ($\mathrm{Schw}_{10}$) when the black hole is small (high temperature).  

To accomplish this, perform a change of coordinates
\begin{eqnarray}
y&=&\sqrt{1-{\rm sech} \left(\rho \,\xi \sqrt{2-\xi^2} \right)}\,, \nonumber\\
x&=&\sqrt{1-\sin \left(\frac{1}{2} \,\rho \left(1-\xi ^2\right)\right)}.
 \label{chartmap}
\end{eqnarray}
This is essentially a Cartesian to polar map. To see this, the transformation $y=\sqrt{1-\rm{sech}(Y)}$ and $x=\sqrt{1-\sin(X/2)}$ maps the $\mathrm dx^2$ and $\mathrm dy^2$ components in the reference metric to $L^2H_2(\mathrm d X^2+\mathrm d Y^2)$, which is conformal to Cartesian coordinates. Finally, $X=\rho\xi\sqrt{2-\xi^2}$ and $Y=\rho(1-\xi^2)$ give the usual Cartesian to polar map with a different angular coordinate. 

In these new coordinates, we rewrite our ansatz
\begin{subequations}\label{eq:ansatzrhoxi}
\begin{align}
       \mathrm ds^2&= -M {f}_1 \frac{\left(\rho^7-\rho_0^7\right)^2}{\left(\rho^7+\rho_0^7\right)^2}  {\mathrm d}t^2+\nonumber\\  
       &\qquad +L^2 H_2  {\biggl [}  \tilde{f}_2  {\mathrm d}\rho^2+\rho^2 {\biggl (} \frac{4\tilde{f}_4 (  {\mathrm d}\xi+ \tilde{f}_6 {\mathrm d}\rho)^2}{2-\xi ^2}+\nonumber\\   
&\qquad+G_1 \xi^2(2-\xi^2){f}_3 {\mathrm d}\Omega_3^2+G_2 \left(1-\xi ^2\right)^2 {f}_5 \, {\mathrm d}\Omega_4^2
{\biggr )}  \biggr ], \\
          C_{(4)}&= L^3 \frac{\xi ^4 \left(2-\xi ^2\right)^2 \rho ^4}{\sqrt{2}} M G_3 {f}_7 \,{\mathrm d}t  \wedge {\mathrm d}S_{(3)}+L^4 W \,dS_{(4)},
\end{align}
\end{subequations}
where $\tilde f_2,\tilde f_4,\tilde f_6$ are new unknown functions of $\rho$ and $\xi$ (the other $f_i$'s transform as scalars).  On the reference metric, these new functions are $\tilde f_2=\tilde f_4=1$ and $\tilde f_6=0$.  The map \eqref{chartmap} uniquely determines $G_1$ and $G_2$, and relates $M$ and $G_3$ directly with $H_1$.   The factor of $(\rho^7-\rho_0^7)^2/(\rho^7+\rho_0^7)^2$ is chosen in anticipation of placing a black hole horizon in the reference metric.  Now set $H_2=(1+\rho_0^7/\rho^7)^{4/7}$.  This is consistent with our requirement that $H_2=1$ at $y=1$ ($\rho\to\infty$).  At small $\rho$ and $\rho_0$, we have $G_1\approx G_2\approx 1$, so if $M\approx1$, the reference metric would approximate $\mathrm{Schw}_{10}$ in isotropic coordinates.  This guides our choice for $H_1$ (and consequently $M$ and $G_3$).  We choose $H_1$ so that $H_1=1$ at $y=1$ ($\rho\to\infty$) and $M$ is positive definite with $M=1$ at $\rho=\rho_0$.  The last requirement also fixes the temperature and ensures regularity on the horizon of the reference metric.  Our specific choice for $H_1$, and explicit expressions for the other known functions are given in the Appendix. 

Our integration domain contains five boundaries: the horizon $\rho=\rho_0$, asymptotic infinity $y=1$ ($\rho\to\infty)$, the $S_3$ axis $y=0$ ($\xi=0$), the $\mathrm{S}^5$ `north' pole $x=1$ ($\xi=1$), and the `south' pole $x=0$. For boundary conditions at infinity $y=1$ ($\rho\to\infty$), we require global $\mathrm{AdS}_5\times \mathrm{S}^5$ asymptotics: $f_1=\ldots=f_5=1$, $f_6=0$, $f_7=1$. The remaining boundary conditions are determined by regularity.  As we have written our functions, we only require $f_i$ and $\tilde f_i$ to remain finite on these boundaries, and more specific boundary conditions can be found through a series expansion of the equations of motion \cite{Dias:2015nua}. 

\begin{figure}[ht]
\centering
\includegraphics[width=.35\textwidth]{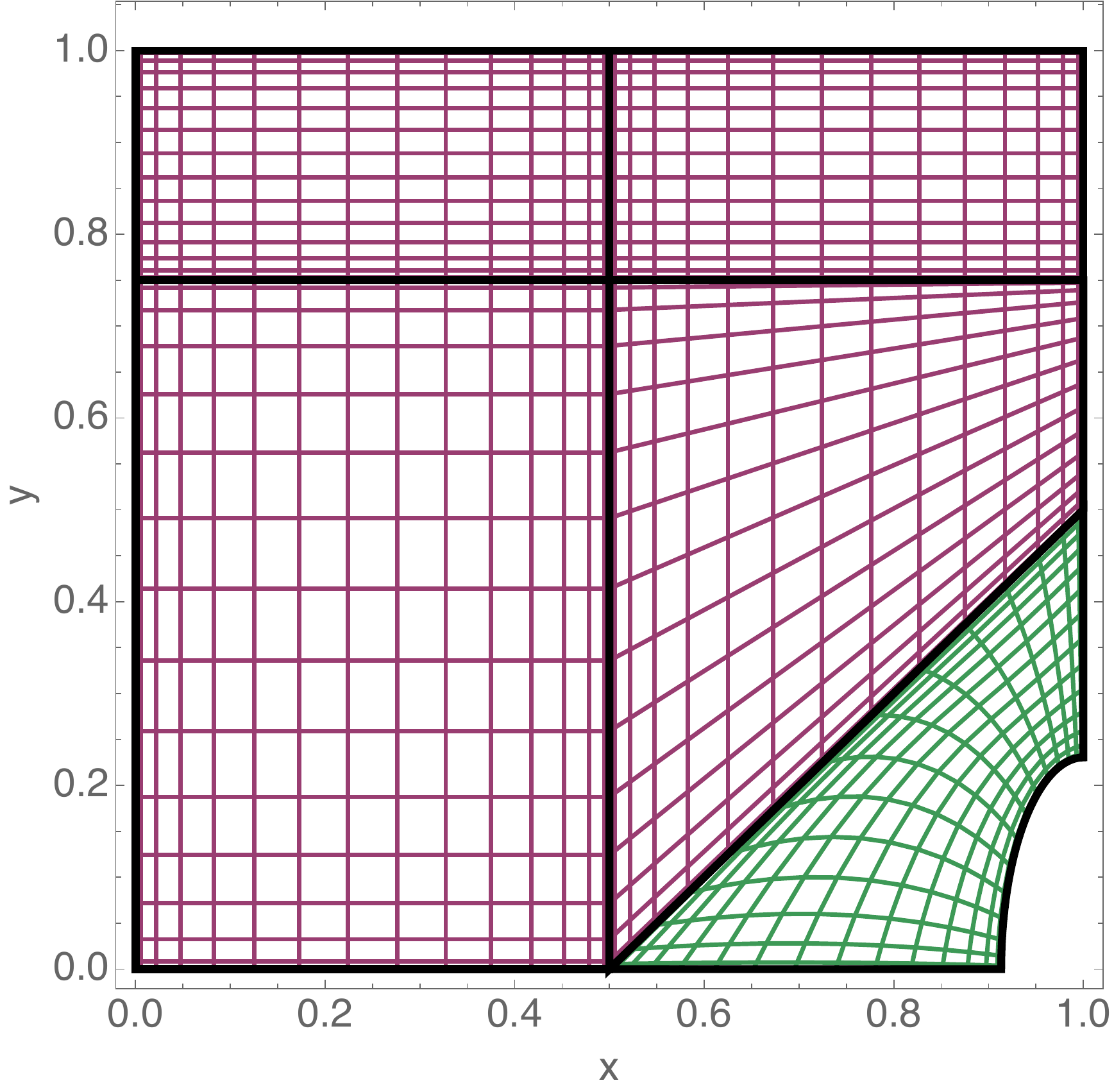}
\caption{Integration domain in $\{x,y\}$ coordinates. The green patch near the horizon is mapped from $\{\rho,\xi\}$ coordinates.}\label{Fig:patches}
\end{figure}

To handle the five boundaries numerically, we divide the integration domain into a number of non-overlapping warped rectangular regions or `patches' as shown in Fig.~\ref{Fig:patches}.  The four patches far away from the horizon use $\{x,y\}$ coordinates, while the remaining patch near the horizon use $\{\rho,\xi\}$ coordinates.  We must supplement our boundary conditions with additional `patching conditions': the metric $g$ and the form $C_{(4)}$ and their first derivatives must match across patch boundaries.  

We therefore have a boundary value problem for 7 functions in two dimensions. $L$ drops out of the equations of motion, so the system is parametrised by $\rho_0$ which fixes the temperature \footnote{Our ansatz limits $\rho_0<\pi$, but our solutions are well within this bound}. We solve the system with Newton-Raphson using the reference metric and $f_7=1$ at $\rho_0=0.1$ as a first seed.  We use pseudospectral collocation with transfinite interpolation of Chebyshev grids in each patch, and the linear systems are solved by LU decomposition.  All solutions we have found satisfy $\xi^2<10^{-10}$.  See Appendix for convergence tests.

\begin{figure}[ht]
\centering
\includegraphics[width=.4\textwidth]{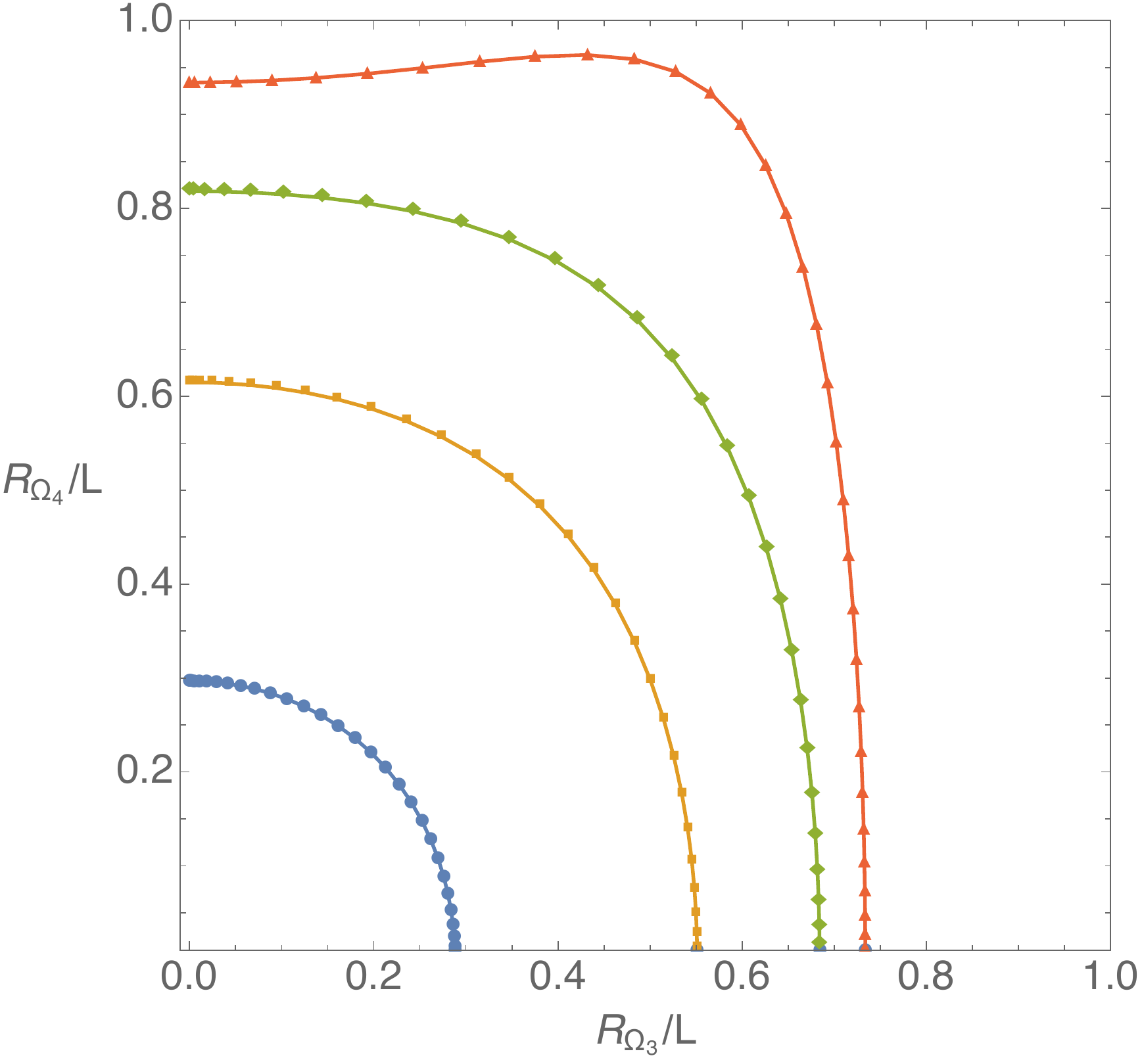}
\caption{Radii of the $S^3$ and $S^4$ along the horizon. From the bottom-left to the top-right $T=\{1.90, 0.945, 0.708, 0.538\}$.}\label{Fig:deformation}
\end{figure}

{\bf 3.~Results --} 
In Fig. \ref{Fig:deformation} we show the radii $R_{\Omega_3}$, $R_{\Omega_4}$ of the geometrically preserved $S^3$ and $S^4$ along the horizon. This curve at small $\rho_0$ (high temperatures) is approximated by $R_{\Omega_3}^2+R_{\Omega_4}^2\approx 2^{4/7}\,\rho_0^2\,L^2$, implying that the horizon is nearly spherical. At larger $\rho_0$ (lower temperatures), the horizon is much more deformed. 

Now we compute thermodynamic quantities.  The temperature $T$ is fixed by $\rho_0$. The entropy $S$ is found by integrating the horizon area. The energy $E$ is computed using the formalism of Kaluza-Klein holography and holographic renormalisation \cite{Skenderis:2006uy,Dias:2015pda,Kim:1985ez,Gunaydin:1984fk,Lee:1998bxa,Lee:1999pj,Arutyunov:1999en,Skenderis:2006di,Skenderis:2007yb} (see \cite{Dias:2015pda} and Appendix for details).  The AdS/CFT dictionary relates the 10 and 5 dimensional Newton constants to the number of colours $N$ of $\mathcal N=4$ SYM via $G_{10}=\frac{\pi^4}{2}\frac{L^8}{N^2}$ and $G_5=\frac{G_{10}}{\pi^3 L^5}$. These yield the expressions
\begin{subequations}\label{eq:thermo}
\begin{align}
        TL&=\frac{7}{2^{16/7} \pi}\,\frac{1}{\rho_0}\;, \\  
        \frac{S}{N^2}&= \frac{2^{44/7}\rho_0^8}{3}\times \nonumber\\ 
&\times\int_0^1 d\xi \, \xi^3(2-\xi^2)(1-\xi^2)^4 {f}_3^{3/2}  \tilde{f}_4^{1/2} {f}_5^2 G_1^{3/2}G_2^2{\biggl |}_{\rho=\rho_0},   \\
\frac{EL}{N^2}&= \frac{1}{512}\left(\partial_y^{(4)} f_3-\partial_y^{(4)} f_1\right){\biggl |}_{y=1}\;.
\end{align}
\end{subequations}
These quantities numerically satisfy the first law $\mathrm dE=T \mathrm dS$ to $<0.1\%$ error.  

\begin{figure}[ht]
\centering
\includegraphics[width=.4\textwidth]{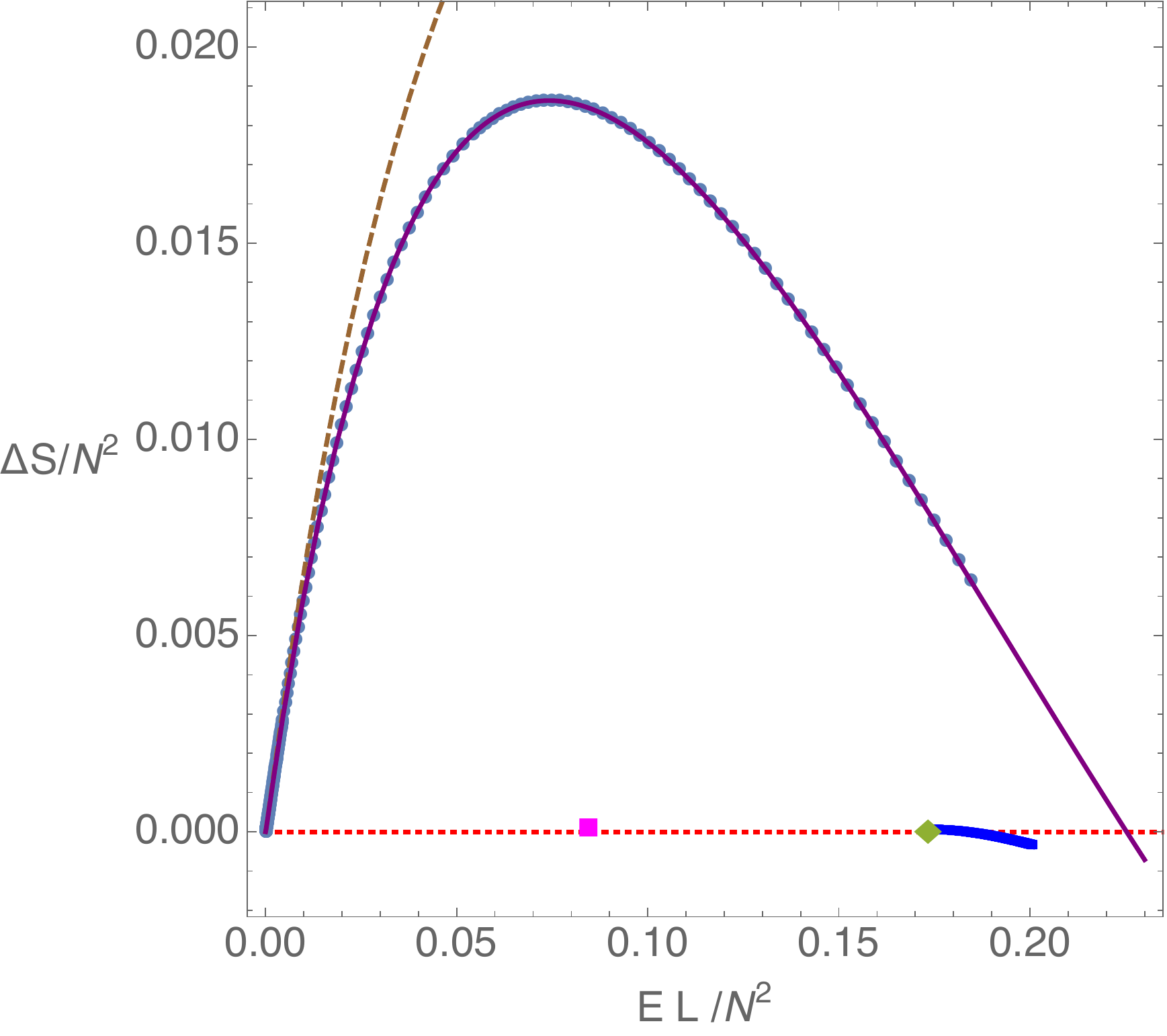}
\caption{Microcanonical phase diagram: entropy with respect to that of $\mathrm{AdSSchw_5}\times \mathrm{S}^5$ vs energy. The dotted red line is the $\mathrm{AdSSchw_5}\times \mathrm{S}^5$ phase, while the blue squares are the $\ell=1$ lumpy black holes. The green diamond and magenta square mark the onset of the $\ell=1$ and $\ell=2$ Gregory-Laflamme instability, respectively. The solid purple curve and its points describe the localised black holes and a fit of its data. The brown dashed line is the lowest-order $\mathrm{Schw}_{10}$ approximation.}\label{Fig:microcanonical}
\end{figure}

In the microcanonical ensemble, the energy is fixed, and the dominant solution maximises the entropy. In Fig. \ref{Fig:microcanonical}, we show $S/N^2$ vs $E L/N^2$ for various competing solutions. The entropy is shown with respect to the entropy of $\mathrm{AdSSchw_5}\times \mathrm{S}^5$ \cite{Hawking:1982dh,Witten:1998zw}. For small energies, the entropy of the localised black hole is well-approximated by that of $\mathrm{Schw_{10}}$ and is larger than that of $\mathrm{AdSSchw_5}\times \mathrm{S}^5$.  

For $E L/N^2\lesssim0.173$, $\mathrm{AdSSchw_5}\times \mathrm{S}^5$ black holes are unstable.  We have found localised black holes for this energy range and determined that they have more entropy than $\mathrm{AdSSchw_5}\times \mathrm{S}^5$, indicating that localised black holes are a plausible endpoint to this instability.

At higher energies, the entropy of localised black holes approaches that of $\mathrm{AdSSchw_5}\times \mathrm{S}^5$, where we believe they will eventually meet in a first-order phase transition. Unfortunately, we were unable to reach this phase transition with our current numerical resources.  An extrapolation of data (see Appendix for details) places the phase transition at $\{E L/N^2,S/N^2\} \approx \{0.225,0.374\}$.

\begin{figure}[ht]
\centering
\includegraphics[width=.4\textwidth]{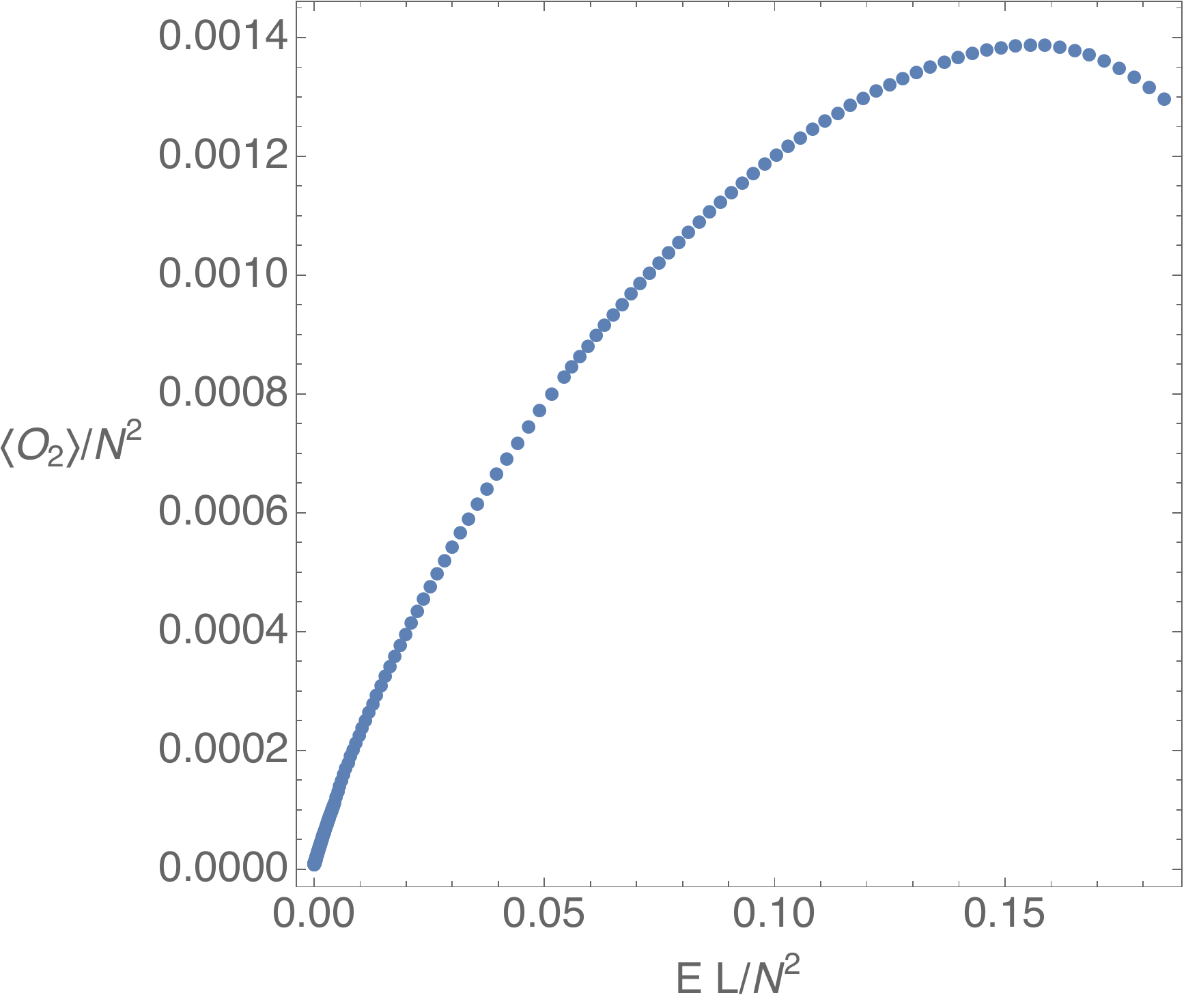}
\caption{Dimension 2 scalar condensate vs energy. 
}\label{fig:expectation}
\end{figure}

The localised black holes dominate the microcanonical ensemble at low energies, but do not respect the full asymptotic $SO(6)$ symmetry of the $\mathrm{S}^5$.  This is dual to a spontaneous symmetry breaking of the $SO(6)$ R-symmetry of the scalar sector of $\mathcal{N}=4$ SYM down to $SO(5)$. This results in a condensation of an infinite tower of scalar operators with increasing conformal dimension. The lowest conformal dimension is 2, and the associated scalar operator can be written as
\begin{equation}\label{eq:operator}
\mathcal O_2 = \frac{2}{g^2_{\mathrm{YM}}}\sqrt{\frac{5}{3}}\mathrm{Tr}\left[(X^1)^2-\frac{1}{5}\Big((X^2)^2+\ldots+(X^6)^2\Big)\right]\;,
\end{equation}
where $X^i$ the are the six real scalars of $\mathcal N=4$ SYM in the vector representation of $SO(6)$ and $g_{\mathrm{YM}}$ is the coupling constant (see e.g. \cite{D'Hoker:2002aw} for the action of $\mathcal N=4$ SYM). The expectation value $\langle \mathcal{O}_2\rangle$ in the broken phase can be found from the supergravity solution through the formalism of Kaluza-Klein holography \cite{Skenderis:2006uy,Dias:2015pda,Kim:1985ez,Gunaydin:1984fk,Lee:1998bxa,Lee:1999pj,Arutyunov:1999en,Skenderis:2006di,Skenderis:2007yb} (see Appendix for details).  We show $\langle \mathcal{O}_2\rangle$ for a range of energies in Fig.~\ref{fig:expectation}.  Because the symmetry breaking transition is first order, $\langle \mathcal{O}_2\rangle$ will have a nonzero value at the phase transition. 

In the canonical ensemble, the temperature is fixed and the solution with lowest Helmholtz free energy $F=E-TS$ dominates. In this ensemble, there is a first order phase transition at the Hawking-Page point $\{F L/N^2,TL\}=\{0, 3/(2\pi)\}$ between large $\mathrm{AdSSchw_5}\times \mathrm{S}^5$ black holes at higher temperatures and thermal $\mathrm{AdS_5}\times \mathrm{S}^5$ at lower. All other known solutions, including localised black holes never dominate the canonical ensemble. (See the Appendix for a phase diagram.)
 
{\bf 4.~Discussion --} 
To summarise, we have numerically constructed asymptotically global $\mathrm{AdS}_5\times \mathrm{S}^5$ localised black holes in type IIB supergravity. These black holes are topologically $\mathrm S^8$ and are more entropic than any other known solution at low energies. At higher energies near $E L/N^2\approx 0.255$, there is a first order phase transition to $\mathrm{AdSSchw}_5\times \mathrm{S}^5$ black holes. By the AdS/CFT correspondence, these localised black holes are dual to a spontaneously broken thermal state of $\mathcal N=4$ super Yang Mills with large $N$ gauge group and large t'Hooft coupling. The scalar sector with the broken symmetry contains a dimension-$2$ operator with an expectation value shown in Fig.~\ref{fig:expectation} and preserves a $SO(5)$ subgroup of $SO(6)$. 

Since lattice simulations of field theories with holographic duals rely on finite temperature, numerical tests of AdS/CFT on both sides of the duality have been restricted to the canonical ensemble \cite{Catterall:2007fp,Anagnostopoulos:2007fw,Catterall:2008yz,Hanada:2008ez,Catterall:2009xn,Hanada:2011fq,Hanada882}. However, there has been recent progress in understanding first order phase transitions in several ensembles \cite{PhysRevLett.68.9, PhysRevLett.71.211, PhysRevE.64.056101}. We emphasise that such field theory calculations on $\mathcal{N}=4$ SYM, at large t'Hooft coupling and large $N$, should reproduce both Fig.~\ref{Fig:microcanonical} and Fig.~\ref{fig:expectation}.

The completion of the phase diagram in Fig.~\ref{Fig:microcanonical} can be conjectured from other systems with Gregory-Laflamme instabilities \cite{Harmark:2002tr,Kol:2002xz,Kudoh:2004hs,Harmark:2007md,Headrick:2009pv,Dias:2014cia,Emparan:2014pra} (see reviews \cite{Harmark:2007md,Dias:2015nua}).  There is a family of `lumpy' black holes \cite{Dias:2015pda} (blue squares in Fig.~\ref{Fig:microcanonical}) connected to the onset of the Gregory-Laflamme instability (green diamond in Fig.~\ref{Fig:microcanonical}).  Lumpy black holes have horizon topology $S^3\times \mathrm{S}^5$, but have nontrivial deformations along the $\mathrm{S}^5$. We expect the localised black holes to meet with the lumpy black holes in the space of solutions. For this to happen without violating the first law, there must be a cusp somewhere in the $S/N^2$ vs $EL/N^2$ curve.  There must also be a topological transition point, which would be a solution containing a naked curvature singularity.  Analogous systems with Gregory-Laflamme instabilities suggest that this topological transition point is closer to the lumpy black hole side of the curve. That is, that the cusp would be a topologically $\mathrm S^8$ black hole. 

Let us now comment on dynamical evolution.  Entropy arguments suggest that the evolution of unstable $\mathrm{AdSSchw}_5\times \mathrm{S}^5$ black holes would proceed towards the most dominant solution, which are the localised $\mathrm S^8$ black holes.  This entails a violation of cosmic censorship, much like in the evolution of the black string \cite{Lehner:2010pn} or black ring \cite{Figueras:2015hkb}. Whether or not the evolution proceeds in this way, and the implications for $\mathcal N=4$ SYM if cosmic censorship is violated remain important open problems. Interestingly, there is a range of energies $0.173\lesssim EL/N^2 \lesssim 0.225$ where $\mathrm{AdSSchw}_5\times \mathrm{S}^5$ is subdominant in entropy but nevertheless dynamically stable. In the field theory, this means that the time scale for spontaneous symmetry breaking at these energies is exponentially suppressed compared to those at lower energies.

Many localised solutions dual to $\mathcal N=4$ SYM states remain to be studied. In global $\mathrm{AdS_5\times \mathrm{S}^5}$, there are localised solutions that break more symmetries, but these are likely less entropic than the ones preserving $SO(5)$.  There are other localised solutions arising from higher harmonics of the Gregory-Laflamme instability.  In particular, the $\ell=2$ mode (whose onset is shown in Figs. \ref{Fig:microcanonical}) leads to double $\mathrm S^8$ black holes and $S^4\times S^4$ `black belts' \cite{Dias:2015pda}.  However, these require delicate balancing of forces and are likely unstable.  Rotational effects remain largely unexplored except for the onset of the Gregory-Laflamme instability for equal spin black holes \cite{Dias:2015nua}. Beyond global $\mathrm{AdS_5\times \mathrm{S}^5}$, there is freedom to choose a different gauge theory background than one conformal to $\mathbb R^{(t)}\times S^3$. This can yield novel physics like plasma balls and boundary black holes (see \cite{Marolf:2013ioa} for a review), but none of these studies have included the effects of localisation.\newpage

\vskip .2cm
\centerline{\bf Acknowledgements}
\vskip .2cm
O.D. acknowledges financial support from the STFC Ernest Rutherford grants ST/K005391/1 and ST/M004147/1. B.W. is supported by European Research Council grant no. ERC-2011-StG 279363-HiDGR. The authors thankfully acknowledge the computer resources, technical expertise and assistance provided by CENTRA/IST. Computations were performed at the cluster ``Baltasar-Sete-S\'ois'' and supported by the H2020 ERC Consolidator Grant ``Matter and strong field gravity: New frontiers in Einstein's theory'' grant agreement no. MaGRaTh-646597.

\onecolumngrid  \vspace{1cm} 
\begin{center}  
{\Large\bf Appendix} 
\end{center} 
\appendix

\section{Ansatz Functions}
Here, we give the known functions in our ansatz \eqref{eq:ansatzxy} and \eqref{eq:ansatzrhoxi} in full.  These are
\begin{subequations}
\begin{align}
	G_1&={\rm sinc}\left(i \,\rho\,\xi  \sqrt{2-\xi ^2} \right)^2\;,\\
	G_2&={\rm sinc}{\biggl (}\rho \left(1-\xi ^2\right) {\biggr )}^2\;,\\
        H_1&=\frac{E_{-}^2}{E_{+}^2}\left(y^2\left(2-y^2\right)\frac{E_{-}^2}{E_{+}^2}+\left(1-y^2\right)^2\right)\;,\\
        H_2&=\left(1+\frac{\rho_0^7}{\rho^7}\right)^{4/7}=\left(1+\frac{\rho_0^7}{\left({\rm arcsech}\left(1-y^2\right)^2+4\,{\rm arcsin}\left(1-x^2\right)^2\right)^{7/2}}\right)^{4/7}\;,\\
        M&=
1+\frac{\left(\rho_0^7-\rho^7\right)^2}{\left(\rho^7+\rho_0^7\right)^2}\sinh^2\left(\rho\,\xi \sqrt{2-\xi ^2} \right)\;,
\end{align}
\end{subequations}

where
\begin{equation}
E_{\pm}=\rho_0^7 \pm \left(4\,{\rm arcsin}\left(1-x^2\right)^2+{\rm arcsech}\left(1-y^2\right)^2\right)^{7/2}\;.
\end{equation}
and the function ${\rm sinc} z=\frac{\sin z}{z}$ for $z\neq 0$ and ${\rm sinc} z=1$ for $z=0$. 

The relationship between $\{x,y\}$ and $\{\rho,\xi\}$ coordinate systems is given by
\begin{subequations}
\begin{align}
	y&=\sqrt{1-{\rm sech} \left(\rho \,\xi \sqrt{2-\xi^2} \right)}\,,\qquad\qquad x=\sqrt{1-\sin \left(\frac{1}{2} \,\rho \left(1-\xi ^2\right)\right)}\;,\\
	\rho&=\sqrt{{\rm arcsech}\left(1-y^2\right)^2+4\,{\rm arcsin}\left(1-x^2\right)^2}\,,\qquad\qquad \xi=\sqrt{1-\frac{2 {\rm arcsin}\left(1-x^2\right)}{\sqrt{{\rm arcsech}\left(1-y^2\right)^2+4\,{\rm arcsin}\left(1-x^2\right)^2}}}\;.
\end{align}
\end{subequations}

\section{Holographic Stress Tensor and Kaluza-Klein holography}
To find the energy $E$, the holographic stress tensor $T_{ij}$ and the expectation value of holographic dual operators we need the asymptotic Taylor expansion of the fields around $y=1$ up to order $\mathcal{O}\left(1-y\right)^4$. This is given by 

\begin{subequations}
\begin{eqnarray}\label{asymptotics}
f_1{\bigl |}_{y=1}&=& 1+ \sqrt{\frac{5}{6}} \beta_2  (1-y)^2 Y_2(x)
-\frac{1}{6} \sqrt{\frac{5}{2}} (1-y)^3 \left(3 \gamma_3   Y_3(x)+2 \sqrt{3} \beta_2  Y_2(x)\right)
\nonumber\\
&& +\frac{1}{192} (1-y)^4 {\biggl [}
Y_0(x) \left(\left(40 \beta_2^2 +5 \beta_2 +12 (16 \delta_0 +\delta_4-192)\right)-2304\right)\nonumber\\
&&
+4 \left(2 \sqrt{30} \beta_2 (8 \beta_2+17) Y_2(x)+\left(18 \sqrt{10} \gamma_3Y_3(x)+5 \sqrt{7} \left(8 \beta_2^2+\beta_2-4 \delta_4\right) Y_4(x)\right)\right)
{\biggr] } +\mathcal{O}\left(1-y\right)^5, \\
f_2{\bigl |}_{y=1}&=& 1+ \sqrt{\frac{5}{6}} \beta_2  (1-y)^2 Y_2(x)
-\frac{1}{6} \sqrt{\frac{5}{2}} (1-y)^3 \left(3 \gamma_3    Y_3(x)+2 \sqrt{3} \beta_2  Y_2(x)\right)\nonumber\\
&&+
\frac{ (1-y)^4}{96} {\biggl [} 85 \beta_2^2  Y_0(x)+4 \sqrt{30} (8 \beta_2 +17) \beta_2  Y_2(x)\nonumber\\
&&+2 \left(18 \sqrt{10} \gamma_3  Y_3(x)+5 \sqrt{7} \left(8 \beta_2^2 +\beta_2 -4 \delta_4\right) Y_4(x)\right) {\biggr ]} 
+\mathcal{O}\left(1-y\right)^5, \\
f_3{\bigl |}_{y=1}&=& 1+ \sqrt{\frac{5}{6}} \beta_2  (1-y)^2 Y_2(x)
-\frac{1}{6} \sqrt{\frac{5}{2}} (1-y)^3 \left(3 \gamma_3  Y_3(x)+2 \sqrt{3} \beta_2  Y_2(x)\right)\nonumber\\
&&+
\frac{1}{576} (1-y)^4 
{\biggl [}  Y_0(x) (2304-(5 \beta_2 +12 (16 \delta_0 +\delta_4-192)))+\nonumber\\
&&
12 \left(2 \sqrt{30} \beta_2  (8 \beta_2 +17) Y_2(x)+\left(18 \sqrt{10} \gamma_3  Y_3(x)+5 \sqrt{7} \left(8 \beta_2^2 +\beta_2 -4 \delta_4\right) Y_4(x)\right)\right) {\biggr ]} 
+\mathcal{O}\left(1-y\right)^5, \\
f_4{\bigl |}_{y=1}&=& 1+\frac{1}{2} \sqrt{\frac{5}{6}} \beta_2  (1-y)^2 \left(45 Y^x_{x\,(2)}(x)-\sqrt{30} Y_0(x)\right) \nonumber\\
&&
-\frac{1}{240} (1-y)^3 {\biggl [}  900 \sqrt{30} \beta_2  Y^x_{x\,(2)}(x)+525 \sqrt{10} \gamma_3  Y^x_{x\,(3)}(x)-600 \beta_2  Y_0(x)+512 \sqrt{5}\gamma_1 Y_1(x)-300 \gamma_3  Y_1(x) {\biggr ]}\nonumber\\
&& +(1-y)^4 
{\biggl [}  \frac{25}{1728 \sqrt{7}} \left(8 \beta_2^2 +\beta_2 -4 \delta_4\right) \left(30 \sqrt{210} Y^x_{x\,(2)}(x)+280 Y^x_{x\,(4)}(x)-21 \sqrt{7} Y_0(x)\right) \nonumber\\
&&+\frac{5}{6} \beta_2^2  \left(4 \sqrt{\frac{6}{5}} Y^x_{x\,(2)}(x)-\frac{4}{3} \sqrt{7} Y^x_{x\,(4)}(x)-\frac{7 Y_0(x)}{16}\right)-\frac{17}{8} \sqrt{\frac{5}{6}} \beta_2  \left(\sqrt{30} Y_0(x)-45 Y^x_{x\,(2)}(x)\right)\nonumber\\
&&+\frac{15}{32} \gamma_3  \left(7 \sqrt{10} Y^x_{x\,(3)}(x)-4 Y_1(x)\right)+\frac{16\gamma_1 Y_1(x)}{\sqrt{5}} {\biggr ]} 
+\mathcal{O}\left(1-y\right)^5, \\
f_5{\bigl |}_{y=1}&=& 1-\frac{1}{2} \sqrt{\frac{5}{6}} \beta_2  (1-y)^2 \left(180 Y^\Omega_{\Omega\,(2)}(x)+\sqrt{30} Y_0(x)\right) \nonumber\\
&&
+\frac{1}{60} (1-y)^3{\biggl [} 75 \sqrt{10} \left(12 \sqrt{3} \beta_2  Y^\Omega_{\Omega\,(2)}(x)+7 \gamma_3  Y^\Omega_{\Omega\,(3)}(x)\right)+150 \beta_2  Y_0(x)+\left(75 \gamma_3 -128 \sqrt{5}\gamma_1\right) Y_1(x)  {\biggr ]} \nonumber\\
&&
-\frac{(1-y)^4}{8640}
{\biggl [}  75 \left(7 \left(46 \beta_2^2 +5 \beta_2 -20 \delta_4\right)+1224 \beta_2 \right) Y_0(x)
-216 \left(128 \sqrt{5}\gamma_1-75 \gamma_3 \right) Y_1(x)\nonumber\\
&&+120 \sqrt{30} \left(1048 \beta_2^2 +5 (918+25) \beta_2 -500 \delta_4\right) Y^\Omega_{\Omega\,(2)}(x)\nonumber\\
&&+200 \left(567 \sqrt{10} \gamma_3  Y^\Omega_{\Omega\,(3)}(x)+4 \sqrt{7} \left(8 \beta_2^2 +25 \beta_2 -100 \delta_4\right) Y^\Omega_{\Omega\,(4)}(x)\right){\biggr ]} 
+\mathcal{O}\left(1-y\right)^5, \\
f_6{\bigl |}_{y=1}&=& 
\frac{1}{8} \sqrt{\frac{5}{2}} \beta_2  \left(2-x^2\right) (1-y) S_x^{(2)}(x)
+\left(2-x^2\right)(1-y)^2{\biggl [} \gamma_1 S_x^{(1)}(x)-\frac{1}{128} \sqrt{10} \left(12 \beta_2  S_x^{(2)}(x)+\sqrt{21} \gamma_3   S_x^{(3)}(x)\right) {\biggr ]} \nonumber\\
&&
-\left(2-x^2\right) (1-y)^3{\biggl [} 2\gamma_1 S_x^{(1)}(x)-\frac{1}{48}\sqrt{\frac{5}{2}}\beta_2  (8 \beta_2 +51) S_x^{(2)}(x)\nonumber\\
&&
-\frac{1}{288} \sqrt{\frac{7}{2}}\left(9 \sqrt{15} \gamma_3  S_x^{(3)}(x)+5 \left(14 \beta_2^2 +\beta_2 -4 \delta_4\right) S_x^{(4)}(x)\right) {\biggr ]} 
+\mathcal{O}\left(1-y\right)^4, \\
f_7{\bigl |}_{y=1}&=& 1+ \sqrt{\frac{10}{3}} \beta_2  (1-y)^2 Y_2(x)
-\frac{1}{3} \sqrt{\frac{5}{2}} (1-y)^3 \left(3 \gamma_3    Y_3(x)+2 \sqrt{3} \beta_2  Y_2(x)\right)\nonumber\\
&&
+\frac{1}{360} (1-y)^4
{\biggl [}  360 \tilde{\delta}_0 Y_0(x)+2 \sqrt{30} \beta_2  (128 \beta_2 +495) Y_2(x)\nonumber\\
&&
+15 \left(18 \sqrt{10} \gamma_3  Y_3(x)+\sqrt{7} \left(46 \beta_2^2 +5 \beta_2 -20 \delta_4\right) Y_4(x)\right) {\biggr ]} 
+\mathcal{O}\left(1-y\right)^5, 
\end{eqnarray}
\end{subequations}
where $Y_\ell(x)$ , with $\ell=0,1,2,\cdots$ are the (regular) scalar harmonics of S$^5$ given by
\begin{equation}
Y_\ell(x)=\frac{\sqrt{3 \pi } 2^{\frac{1}{2} (-\ell -5)} \sqrt{(\ell +2) (\ell +3)}}{x^{3/2} \left(1-x^2\right)^{3/2} \left(2-x^2\right)^{3/4}}\,P_{\ell +\frac{3}{2}}^{-\frac{3}{2}}\left(-2 x^4+4 x^2-1\right),
\end{equation}
$S_x^{(\ell)}(x)$ is the first component of the scalar derived vector harmonic $S_a^{(\ell)}$, and $Y^x_{x\,(\ell)}(x)$ and $Y^\Omega_{\Omega\,(\ell)}(x)$ are components of the scalar derived tensor harmonic $Y^a_{b\,(\ell)}(x)$ defined as
\begin{equation}
S_a^{(\ell)}=-\frac{1}{\sqrt{\ell(\ell+4)}}D_a Y_\ell, \qquad\qquad Y^a_{b\,(\ell)}(x)=\frac{1}{\ell (\ell +4)}\,D^a D_b Y_\ell+\frac{1}{5}\,\gamma ^a{}_b Y_\ell,
\end{equation}
with $\gamma_{ab}$ being the metric of the S$^5$.

In the above expansion we have already imposed the boundary conditions. The harmonic coefficients depend on six undetermined constants $\{ \beta_2, \gamma_1,\gamma_3, \delta_0, \tilde{\delta}_0, \delta_4\}$. Two of these are gauge modes: $\gamma_1$ can be eliminated using diffeomorphisms while $\tilde{\delta}_0$ can be removed by a gauge transformation of $C_{(4)}$. Accordingly, no physical observable depends on $\gamma_1,\tilde{\delta}_0$. On the other hand, $\{ \beta_2, \gamma_3, \delta_0, \delta_4\}$ are determined only after solving the entire boundary value problem. They vanish for $\mathrm{AdSSchw}_5\times \mathrm{S}^5$ and other solutions that preserve the  full symmetries of the $\mathrm{S}^5$, but not for solutions that break these symmetries. The reader interested on more details about these constants and their relation with the conformal dimensions of the holographic dual operators of the system is invited to read the detailed discussion in Appendix A.5 of \cite{Dias:2015pda}. 

With the above asymptotic expansion of the fields at the holographic boundary, we can compute the holographic stress tensor, the associated energy and expectation values of dual operators, which depend on the constants $\{ \beta_2, \gamma_3, \delta_0, \delta_4\}$.  We can do so using the formalism of Kaluza-Klein holography and holographic renormalisation \cite{Skenderis:2006uy,Dias:2015pda} (see also \cite{Kim:1985ez,Gunaydin:1984fk,Lee:1998bxa,Lee:1999pj,Arutyunov:1999en,Skenderis:2006di,Skenderis:2007yb}). In particular, a detailed discussion of the formalism and expectation value computations for a system like ours that breaks the $SO(6)$ symmetry group of S$^5$ down to $SO(5)$ can be found in Appendix A of \cite{Dias:2015pda}.

The expectation value of holographic stress tensor of our solutions is
\begin{equation} \label{eq:holoT}
\langle T_{ij} \rangle =  \frac{N^2}{2 \pi ^2} \left[ \frac{3}{16}+\frac{3}{4}-\frac{1}{3072} {\biggl(}30 \beta_2^2+5 \beta_2+12 (16 \delta_{0}+\delta_4-192){\biggr)} \right]\, {\rm diag} \left\{ 1, \frac{1}{3}\,\eta_{\hat{i} \hat{j}} \right\},
\end{equation}
with $\eta_{\hat{i}\hat{j}}$ being the metric components of a unit radius S$^3$. This holographic stress tensor is conserved, $ \nabla_i \langle T^{ij} \rangle =0$, and traceless, $\langle T^{\:i}_i \rangle=0$. 

As usual in holographic renormalization, we can now use \eqref{eq:holoT} to read the energy of the solution of our black holes (measured with respect to the global AdS$_5\times$S$^5$ solution):
\begin{equation} \label{eq:energy}
E=\frac{N^2}{3072}{\biggl [}4608 - {\biggl (}5\, \beta_2+30 \,\beta_2^2+12\left(16 \,\delta_{0}+\delta_4\right){\biggr)}{\biggr ]},
\end{equation}
which, using \eqref{asymptotics}, can be rewritten as \eqref{eq:thermo}, namely $\frac{EL}{N^2}= \frac{1}{512}\left(\partial_y^{(4)} f_3-\partial_y^{(4)} f_1\right){\bigl |}_{y=1}$.  Note that we have a static solution with a boundary metric that contains a symmetric $S^3$.  By symmetry and the tracelessness of the stress tensor, all components of the stress tensor can be written in terms of the energy.  

Kaluza-Klein holography also allows us to compute the expectation values of the scalar operators that condensate 
on the boundary theory when the $SO(6)$ R-symmetry of the scalar sector of $\mathcal{N}=4$ SYM is spontaneously broken. There is an infinite tower of such operators but one of them has the lowest conformal dimension, $\Delta=2$. The expectation value of this operator ${\cal O}_{2} $ is 
\begin{equation}\label{eq:vevS2} 
\left \langle {\cal O}_{2} \right\rangle =-\frac{N^2}{\pi^2}\,\frac{1}{8} \sqrt{\frac{5}{3}}  \,\beta_2,
\end{equation}
and this is the expectation value that we that we show in Fig.~\ref{fig:expectation}. 

We can relate this expectation value to quantities in $\mathcal N=4$ SYM.  Recall that the bosonic sector of $\mathcal N=4$ SYM contains six scalars $X^i$, here in the 6-dimensional rank-2 antisymmetric tensor representation of $SU(4)$ (recall that the groups $SU(4)$ and $SO(6)$ have isomorphic Lie algebras).  There is also a spin-1 gauge field $A_\mu$. The Lagrangian of this sector of the theory is given by  \cite{D'Hoker:2002aw}
\begin{equation}\label{eq:N4SYM} 
\mathcal{L}^{\mathrm{(bosonic)}}_{\mathrm{SYM}}={\rm Tr}\left( -\frac{1}{2 \, g_{\rm YM}^2} F_{\mu\nu}F^{\mu\nu}-\sum_i D_\mu X^i D^\mu X^i +\frac{1}{2}\,g_{\rm YM}^2\sum_{i,j} \left[ X^i,X^j \right]^2 \right),
\end{equation}
where $D_\mu X=\partial_\mu X+{\rm i} \left[ A_\mu, X \right] $ is the gauge covariant derivative 
of the theory, $F_{\mu\nu}=\partial_\mu A_\nu-\partial_\nu A_\mu$ and $g_{\rm YM}$ is the dimensionless Yang-Mills gauge coupling. In terms of these fields, the expectation value in \eqref{eq:vevS2} can be written as \eqref{eq:operator} \cite{Witten:1998qj}, though \eqref{eq:operator} uses the vector representation of $SO(6)$.

There is a technical detail that we have not mentioned in the text. In the ansatz \eqref{eq:ansatzxy} and \eqref{eq:ansatzrhoxi}, there is a cross term which can in the $\{x,y\}$ coordinates be generally written schematically as $\sim(1-y^2)^p f_6\mathrm dx \mathrm dy$, for some power $p$.  Note that on the reference metric, we have $f_6=0$, so the reference metric is unaffected by $p$.  However, the boundary condition $f_6=0$ at infinity $(y=1)$ \emph{is} affected by the power $p$. The choice of $p$ therefore holds physical significance, and our choice of $p=0$ is such that the various operators in the dual field theory are unsourced.  For more details into how this power is determined, we refer the reader to the Appendix A.5 of \cite{Dias:2015pda}.

\section{Phase Diagram in the Canonical Ensemble}

\begin{figure}[ht]
\centering
\includegraphics[width=.45\textwidth]{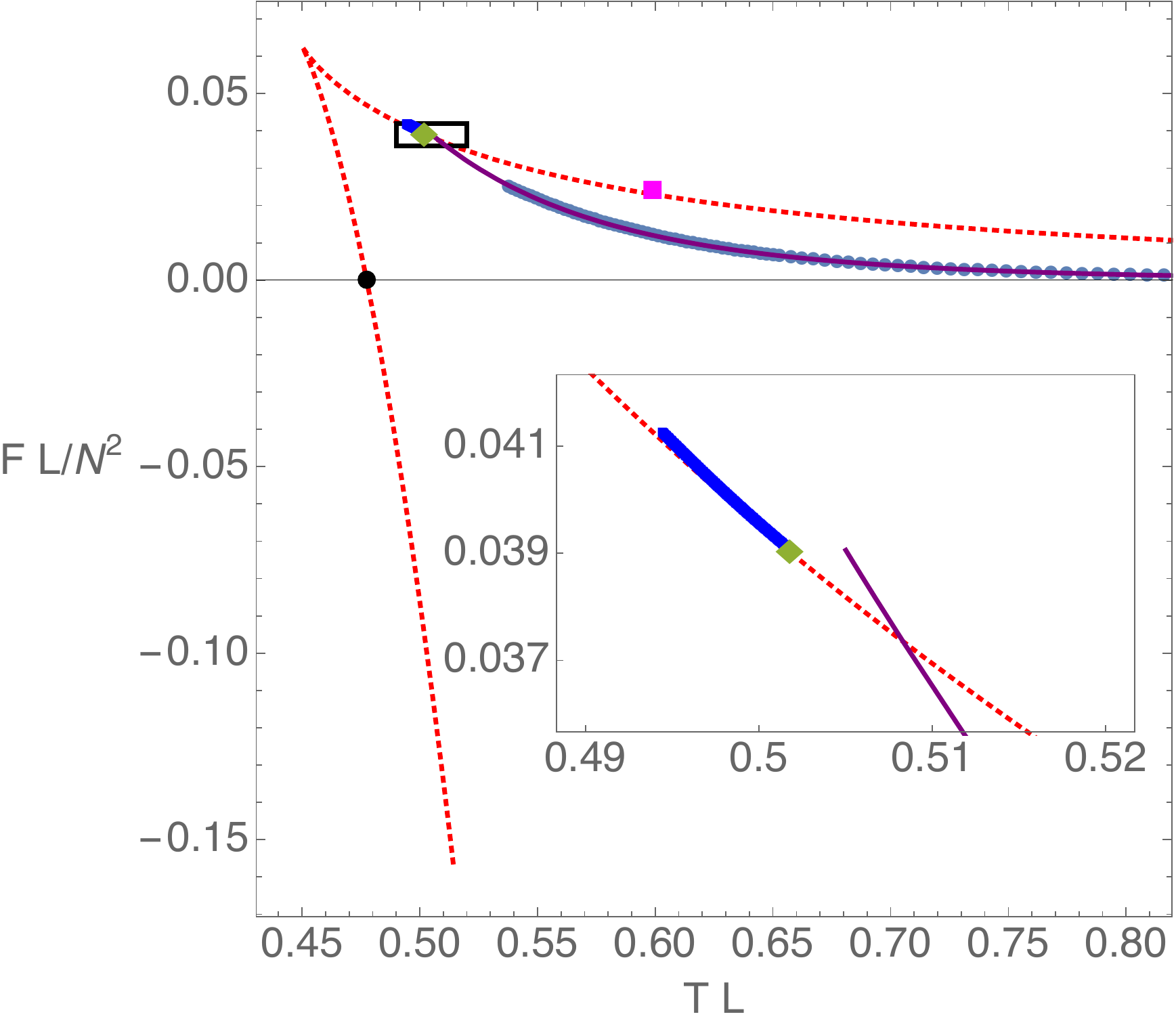}
\caption{Free energy vs Temperature. Same colour scheme as Fig. \ref{Fig:microcanonical}. The black dot marks the Hawking-Page point, and the thin line with $F=0$ represents thermal $\mathrm{AdS}$. 
}\label{Fig:canonical}
\end{figure}

Below we give a phase diagram of our solutions in the canonical ensemble.  In this ensemble, the temperature is fixed and the solution with the lowest free energy dominants. Fig.~\ref{Fig:canonical} shows the free energy $FL/N^2$ versus the temperature $TL$. In this ensemble, there is a first order phase transition at the Hawking-Page point $\{FL/N2,TL\} = \{0,3/(2\pi)\}$ between large black holes at higher temperatures and thermal AdS at lower. All other known solutions are subdominant to these. 

\section{Numerical validation and convergence}
\begin{figure}[ht]
\centering
\includegraphics[width=.9\textwidth]{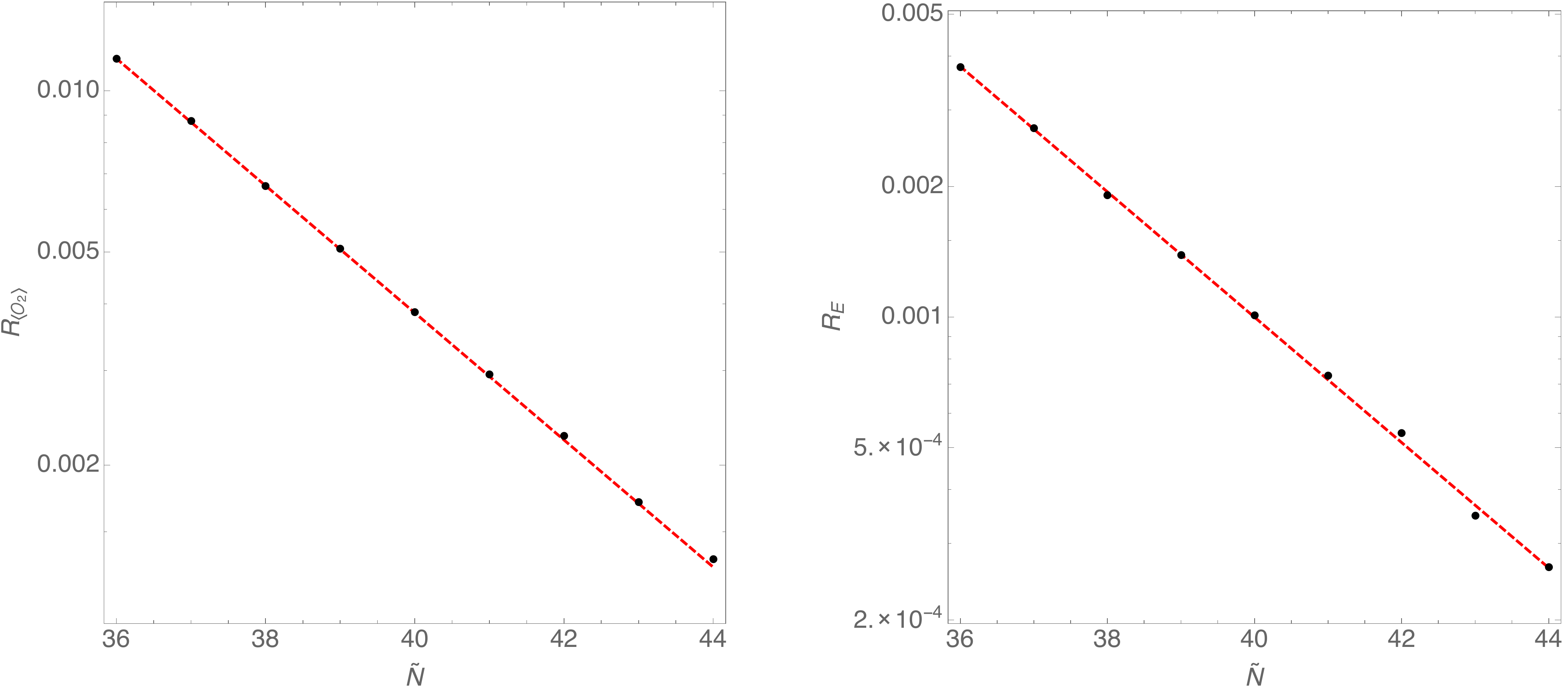}
\caption{Convergence of $\langle \mathcal{O}_2\rangle$ (on the left)  and $E$ (on the right) and  as a function of $\widetilde{N}$ for $\rho_0=17/20$.}
\label{fig:conv}
\end{figure}

In this section, we perform a number of numerical checks. Let us first present convergence tests. Within each patch, we have use a $\widetilde{N}\times \widetilde{N}$ size grid. The convergence of a quantity $Q$ can be shown through the function
\begin{equation}
R_{Q}(\widetilde{N})=\left|1-\frac{Q_{\widetilde{N}}}{Q_{\widetilde{N}+1}}\right|\,,
\end{equation}
which vanishes for large $\widetilde N$ for any converging numerical method. Since we are using pseudospectral collocation, $R_{Q}$ should decrease exponentially in $\widetilde N$ if the solution is sufficiently smooth. Since our reference metric has been adapted for small black holes, it is especially difficult to perform accurate numerics on large black holes. We therefore perform convergence tests for localised black holes with $\rho_0 = 0.85$ which is the largest value we have reached.  In Fig.~\ref{fig:conv} we present convergence tests for the quantities $Q=\langle \mathcal{O}_2\rangle$ and $Q=E$, both of which show exponential convergence.

We can also test our numerics by using proven identities for solutions.  For instance, the energy extracted at infinity via Kaluza-Klein holography needs to be equal to the energy extracted by integrating the first law.  We find that these agree numerically to within $0.1\%$ error. 

The first law, however, does not test all components of the Einstein equation. A more stringent test can be made by investigating Komar identities, similar to the ones presented in \cite{Costa:2014wya}. For every Killing vector $\xi^M$ of our solution we can define an antisymmetric conserved tensor
\begin{equation}
(K_\xi)^{MN}=\nabla^M \xi^N-\frac{1}{12}F_{(5)}^{MNPQR}\xi^U {C_{(4)}}_{PQRU}+\gamma \xi^{[M}F_{(5)}^{N]PQRU}{C_{(4)}}_{PQRU}\,,
\end{equation}
where $F_{(5)} = \mathrm{d}C_{(4)}$ and $\gamma$ is an arbitrary constant. Conservation of this tensor follows from the Einstein equation, $\nabla^M {F_{(5)}}_{MNPQR} = 0$, and from the identities
\begin{equation}
\mathcal{L}_\xi g = \mathcal{L}_\xi C_4=\mathcal{L}_\xi F_5=0\,,\quad\nabla_M\nabla^M\xi_N = -R_{NM}\xi^M\quad\text{and}\quad F_{(5)}=\star F_{(5)}\,,
\end{equation}
where $\mathcal{L}_\xi$ is the Lie derivative along $\xi$.

In the language of differential forms, this means that we have a closed 8-form
\begin{equation}
\mathrm{d}(\star K_\xi)=0\,,
\end{equation}
where
\begin{equation}
K_\xi=\frac{1}{2}(K_\xi)_{MN}\mathrm{d}x^M\wedge \mathrm{d}x^N\,.
\end{equation}

Integrating $\mathrm{d}(\star K_\xi)$ over a $9-$dimensional surface $\Sigma$ of constant time with $y_1 < y < y_2$, we conclude that
\begin{equation}
0=\int_{\Sigma}\mathrm{d}(\star K_\xi)=\int_{\partial_\Sigma}\star K_\xi=\int_{\Gamma(y_1)}\star K_\xi-\int_{\Gamma(y_2)}\star K_\xi+\int_{\mathrm{Dirac}}\star K_\xi\,,
\end{equation}
where we used the fact that the boundary of $\Sigma$ has two disjoint components $\Gamma(y_1)$ and $\Gamma(y_2)$ (with opposite orientations). The last term come from the contribution of a Dirac string, which we will now explain. At infinity, $F_{(5)}=\mathrm{d}C_{(4)}$ contains a term proportional to the $\mathrm{S}^5$ volume form $\mathrm{d}S_{(5)}$, which implies that $C_{(4)}$ cannot be made everywhere regular on the $\mathrm{S}^5$. One can choose $C_{(4)}$ to be regular at the north pole of the $\mathrm{S}^5$, say, but not at the south pole of the $\mathrm{S}^5$. In other words, $C_{(4)}$ has a Dirac string, which we must integrate over. However, we find that for $\gamma=0$, the last term is actually absent. From here onwards, we take $\gamma$ to be zero.

This shows that (for $\gamma=0$) the integral
\begin{equation}
I_\xi(y)=\int_{\Gamma(y)}\star K_\xi
\end{equation}
over the closed surface of constant time and radial coordinate $y$ is independent of the value of $y$. The Komar formula is obtained by equating the integral over the horizon $I_\xi(0)$, which can be written in terms of the entropy and the temperature, to the integral at infinity $I_\xi(1)$, which can be written in terms of the asymptotic quantities of the previous section.
\begin{equation}
\frac{S T}{N^2}=\frac{36 \left(192-8 \delta _0-\delta _4\right)-\beta _2 \left(86 \beta _2+207\right)}{4608}\,.
\end{equation}
Note that in our setup, this identity relates quantities on different patches and different coordinate systems. We have tested this identity, and we find agreement up to $0.1\%$.

Since our numerics were unable to reach the phase transition between localised black holes and $\mathrm{AdSSchw}_5\times \mathrm{S}^5$ black holes, we had to use extrapolation to obtain the location of the phase transition.  We preform this extrapolation by a motivated $\chi^2$ fit to our data on the $\Delta S(E L/N^2)/N^2$ curve (recall $\Delta S$ is the entropy with respect to that of $\mathrm{AdSSchw}_5\times \mathrm{S}^5$). Our fitting function was chosen to be
\begin{equation}
\Delta S_{\mathrm{fit}}(x)/N^2 = \frac{105^{1/7} \pi ^{8/7} x ^{8/7}}{2^{12/7}}\left(1+a_0 x ^{\alpha }+b_0 x ^{\alpha +1}\right)\,.
\end{equation}
The first term in the fitting function was chosen so that it matches the entropy of a ten-dimensional asymptotically flat Schwarzschild black hole at small energies (small $x$). There are a total of three fitting parameters: $a_0$, $b_0$ and $\alpha$. A $\chi^2$ fit yields the values and associated errors $\alpha = 0.190576(1\pm0.0019)$, $a_0 = -0.12688(1\pm0.0011)$ and $b_0=0.187541(1\pm0.0026)$. We can then find the transition point from a simple root-finding algorithm.  This occurs for $E L/N^2 \approx 0.225 (1\pm 0.0027)$, which is the value we quote in the main text. The error obtained for the crossing can be computed by propagating the error associated with each of the parameters in the fit.

To test the sensitivity of this result to other extrapolation methods, we have also performed an order-$p$ polynomial interpolation on the last $p+1$ data points for $p$ from 3 to 10, then extrapolating this polynomial to the phase transition. The largest deviation from $E L/N^2 \approx 0.225$ was under $2\%$.

\bibliography{refs}{}
\end{document}